\documentclass[prl,showpacs,footinbib,twocolumn,final]{revtex4}
\usepackage{graphics,graphicx,color}
\begin{document}

\title{ Scaling of the low temperature dephasing rate in Kondo systems}
\author{F. Mallet$^{1,2}$, J. Ericsson$^{1,2}$, D. Mailly$^{4}$, S. \"{U}nl\"{u}bayir$^{5}$, D. Reuter$^{5}$, A. Melnikov$^{5}$, A.~D.~ÊWieck$^{5}$, T. Micklitz$^{6}$, A. Rosch$^{6}$, T.A. Costi$^{7}$, L. Saminadayar$^{1,2,3}$, and C. B\"auerle$^{1,2}$}
\affiliation{$^{1}$Centre de Recherches sur les Tr\`es Basses
Temp\'eratures, B.P. 166 X, 38042 Grenoble Cedex 09, France}
\affiliation{$^{2}$Institut  N\'{e}el, B.P. 166 X, 38042 Grenoble
Cedex 09, France} \affiliation{$^{3}$Universit\'{e} Joseph Fourier,
B.P. 53, 38041 Grenoble Cedex 09, France}
\affiliation{$^{4}$Laboratoire de Photonique et Nanostructures,
route de Nozay, 91460 Marcoussis, France}
\affiliation{$^{5}$Lehrstuhl f\"{u}r Angewandte
Festk\"{o}rperphysik, Ruhr-Universit\"{a}t, Universit\"atsstr. 150,
44780 Bochum, Germany} \affiliation{$^{6}$Institute for Theoretical
Physics, University of Cologne, 50937 Cologne, Germany}
\affiliation{$^{7}$Institut f\"{u}r Festk\"{o}rperforschung,
Forschungszentrum J\"{u}lich, 52425 J\"{u}lich, Germany}

\date{\today}
\pacs{73.23.-b, 03.65.Bz, 75.20.Hr, 72.70.+m, 73.20.Fz}

\begin{abstract}
We present phase coherence time measurements in
quasi-one-dimensional Ag wires doped with Fe Kondo impurities of
different concentrations $n_s$. Due to the relatively high Kondo
temperature $T_{K}\approx 4.3K$ of this system, we are able to
explore a temperature range from above $T_{K}$ down to below
$0.01\,T_{K}$. We show that the magnetic contribution to the
dephasing rate $\gamma_m$ per impurity is described by a single,
universal curve when plotted as a function of $(T/T_K)$. For
$T>0.1\,T_K$, the dephasing rate is remarkably well described by
recent numerical results for spin $S=1/2$ impurities. At lower
temperature, we observe deviations from this theory. Based on a
comparison with theoretical calculations for $S>1/2$, we discuss
possible explanations for the observed deviations.

\end{abstract}
\maketitle The Kondo effect has been a central theme in solid state
physics for several decades. The fascination it still arouses is due
to the fact that it represents a paradigm of the generic problem of
dynamical impurities in metals~\cite{hewson}. Its most well-known
manifestation is the logarithmic increase of the low temperature
resistivity of metals containing magnetic impurities
\cite{kondo_64}. The phenomenon is accompanied by a progressive
screening of the spin of the impurities by the surrounding
conduction electrons, resulting, in the case of complete screening,
in a Fermi liquid ground state \cite{nozieres_jltp_74}. In the last
decade, a number of experiments have also demonstrated
\cite{mohanty_prl_97,birge_prl_02,schopfer_prl_03,pierre_prb_03}
that Kondo impurities provide an important mechanism for electronic
decoherence at low temperature. Conversely, measurement of the phase
coherence time is a powerful tool to probe the ground state of a
Kondo system.

On the theoretical side, it has been shown very recently
\cite{rosch_prl_06} that the dephasing rate of electrons scattering
from diluted Kondo impurities can be calculated exactly from the
inelastic scattering cross section \cite{zarand_prl_04} by using
renormalization group (NRG) techniques. The calculated full
temperature dependence  of the dephasing rate \cite{rosch_prl_06}
thereby allows for a quantitative comparison between experimental
data and theoretical predictions, bridging the gap between the low
temperature Fermi liquid theory and the \textquotedblleft
high\textquotedblright~temperature Nagaoka-Suhl expansion. Indeed,
recent experiments \cite{bauerle_prl_05} have confirmed qualitative
aspects of these NRG calculations, in particular, the dephasing rate
has been found to be linear with temperature over almost one decade
below $T_{K}$. The relatively low Kondo temperature of the system
used in that work, on the other hand, did not allow to perform
measurements well below $0.1\,T_K$ where the Fermi liquid regime
should appear. In this context, a natural challenge is to
investigate the very low temperature limit of the magnetic
contribution to the dephasing rate $\gamma_m$.

In this Letter we present measurements of the phase coherence time
$\tau_\phi$ of AgFe Kondo wires with different magnetic impurity
concentrations and down to temperatures of $0.01\,T_K$. We show that
the magnetic contribution to the dephasing rate $\gamma_m$ per
magnetic impurity presents a universal scaling when plotted as a
function of $T/T_K$. This dephasing rate is remarkably well
described by the recent numerical result for spin $S=1/2$ impurities
\cite{zarand_prl_04,rosch_prl_06}. On the other hand, at very low
temperatures $(T<0.1\,T_K)$, we observe deviations from this result.

In order to be able to explore the very low temperature limit of
Kondo systems, several severe experimental requirements have to be
met. First of all, a suitable Kondo system with a relatively high
Kondo temperature is needed in order to be able to attain
temperatures of the order of $0.01\,T_K$. At the same time one would
like to avoid that the Kondo maximum of the dephasing rate
\begin{table}[h]
\squeezetable
\begin{tabular}{c c c c c c c c c}
{Sample}&{$l$}&{$R$}&{$\rho$}&{$D$}&{dose}&{$n_{s}$}\\
{}&{($\mu$m)}&{($\Omega$)}&{($\mu\Omega cm$)}&
{(cm$^{2}$/s)}&{($ions/cm^2$)}&{$(ppm)$}\\
\hline
\vspace{-2mm} \\
{AgFe1}&{$745$}&{$877$}&{$0.88$}&{$429$}&{$1.0\cdot 10^{12}$}&{$2.7$}\\
{AgFe2}&{$405$}&{$517$}&{$0.96$}&{$400$}&{$1.0\cdot 10^{13}$}&{$27$}\\
{AgFe3}&{$160$}&{$222$}&{$1.04$}&{$360$}&{$2.5\cdot
10^{13}$}&{$67.5$} \\
{Ag2}&{$765$}&{$1305$}&{$1.28$}&{$295$}&{0}&{$0$}
\end{tabular}
\caption{Sample characteristics: $ l, R, \rho$, $D$ and $n_{s}$
correspond to the length, electrical resistance, resistivity,
diffusion coefficient, and impurity concentration, respectively. All
samples have a width of $w=150\,nm$ and thickness of $t=50\,nm$.}
\label{table}
\end{table}
is masked by the phonon contribution. Secondly, one has to be able
to start with a very pure sample which already shows good agreement
with theory in the absence of magnetic impurities and in the
temperature range of interest. In addition, one would like to be
able to use identical samples and add magnetic impurities with a
very high precision: this is obviously an experimental challenge
which has not been achieved up to date.

\begin{figure}[h]
\includegraphics[width=8cm]{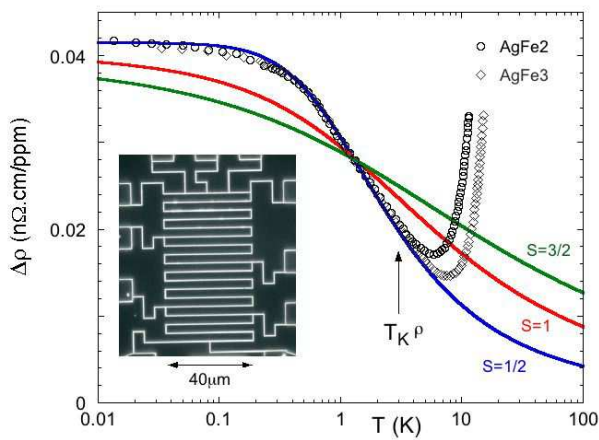}
  \caption{(color online) Resistivity per magnetic impurity concentration as a function
  of temperature for sample AgFe2 and AgFe3. The experimental data have been superposed at low temperatures
   and the electron-electron contribution has been subtracted.
   The solid curves are the NRG results for $S=1/2$ (blue), $S=1$ (red) and $S=3/2$
   (green).}
   \label{R_T}
 \end{figure}

A Kondo system which satisfies the above conditions is Ag/Fe: the
Kondo temperature of this system determined from M\"ossbauer
measurements is around $3\,K$ \cite{rizutto} close to the value we
extract below from resistivity measurements. For this system we can
easily attain temperatures well below $0.1\,T_K$. In addition we are
able to prepare extremely pure silver samples \cite{pierre_prb_03}.
The idea is hence to fabricate several identical quantum wires (see
inset of figure \ref{R_T}) on the same wafer. Subsequently, the
samples are implanted in a very controlled manner with magnetic
impurities via Focused Ion Beam (FIB) technology. The Fe$^{+}$ ion
implantation was done with an energy of 100 keV and the implanted
ion dose is determined directly via the ion current
\cite{comment_SRIM}. In table \ref{table}, the geometrical and
electrical parameters as well as the implanted ion concentrations of
our samples are summarized.

We first determine the Kondo temperature via standard resistivity
measurements. The electrical resistance of a quasi one-dimensional
metallic wire containing magnetic impurities at low temperature is
given by
\begin{equation}
R(T)=  R_0 + \alpha/\sqrt{T} + n_s*f(T/T_K) \label{eq_R_T}
\end{equation}
where $R_0$ is the residual resistance and $n_s$ is the magnetic
impurity concentration. The second term corresponds to the
electron-electron interaction term \cite{AAK_82,gilles_book} and the
third term to the Kondo contribution. The electron-electron
interaction contribution can be determined independently by
measuring the resistivity of a clean sample (Ag\,2) containing no
magnetic impurities. Fitting the temperature dependence of the
resistance variation to $\Delta R(T)=\alpha_{exp}/\sqrt{T}$ we
obtain $\alpha_{exp}\,=\, 0.081\,\Omega /K^{1/2} $, in good
agreement with the theoretical value $\alpha_{theo}\,= 2R^2/R_K/L
\sqrt{\hbar D/k_B T} = \,0.082\,\Omega /K^{1/2}$.

The Kondo contribution to the resistivity per impurity concentration
is shown in figure \ref{R_T} for sample AgFe\,2 and AgFe\,3. Both
data sets scale nicely with impurity concentration and one observes
typical features of a Kondo system: a logarithmic increase below the
Kondo minimum and a saturation at the lowest temperatures. Fitting
the resistivity data to the NRG results for S=1/2
\cite{costi_94,costi_prl_00} we obtain a Kondo temperature of
$T_K^\rho=3.0\,K\pm0.3K$.

\begin{figure}[h]
\includegraphics[width=7.5cm]{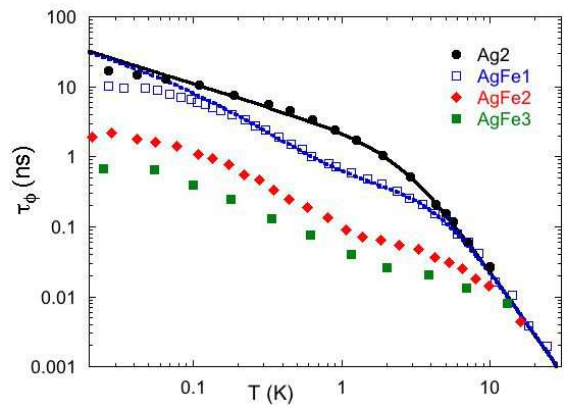}
 \caption{(color online) Phase coherence time of Ag wires containing
 different amounts of Fe impurities. The black solid line corresponds
 to the assumption that only electron-electron and electron-phonon
 interaction contribute to dephasing. The blue dotted line is a
 fit to the experimental data taking into account the
magnetic contribution to the dephasing for S=1/2 impurities.}
 \label{tau_phi}
 \end{figure}

Phase coherence measurements, on the other hand, are much more
sensitive to the presence of magnetic impurities. Here we determine
the phase coherence time $\tau_\phi$ as displayed in
figure~\ref{tau_phi} by fitting the low field magnetoresistance to
standard weak localisation theory \cite{AAK_82,hikami_80}.

The experimentally measured phase coherence time $\tau_{\phi}$ for a
metallic quantum wire containing magnetic impurities can be
described by the following expression \cite{comment_adding}
\begin{equation}
\frac{1}{\tau_\phi} = \frac{1}{\tau_{e-e}} + \frac{1}{\tau_{e-ph}}+
\frac{1}{\tau_{m}} \label{eq_tauphi}
\end{equation}
where
\begin{equation}
\frac{1}{\tau_{e-e}}=a_{theo}\,T^{2/3}=\bigg[\frac{\pi}{\sqrt{2}}\frac{R}{R_K}
\frac{k_B}{\hbar} \frac{\sqrt{D}}{L}\bigg]^{2/3}  T^{2/3}
\end{equation}
corresponds to the electron-electron interaction term
\cite{gilles_book}, $1/\tau_{e-ph}\,=\,b\,T^{3}$ to the
electron-phonon interaction, while $1/\tau_{m}=\gamma_m$ corresponds
to the magnetic contribution to the dephasing. In absence of
magnetic scattering, the temperature dependence of the phase
coherence time is determined by the first two terms in
eq.\ref{eq_tauphi} \cite{AAK_82,comment_adding}. Our data on the
clean wire (sample Ag\,2) follow nicely the expected temperature
dependence down to 40\,mK as shown by the black solid line. From
fitting the data we extract $a_{exp}\,=\,0.42\,ns^{-1}K^{2/3}$ and
$b\,=\,0.04\,ns^{-1}K^{3}$. The value of $a_{exp}$ is somewhat
larger than $a_{theo}=\,0.22\,ns^{-1}K^{2/3}$. At very high
temperatures $(T>5K)$, all samples follow essentially the same
temperature dependence. In this temperature regime the phase
coherence time is mainly determined by electron-phonon scattering.

\begin{figure}[h]
\includegraphics[width=8cm]{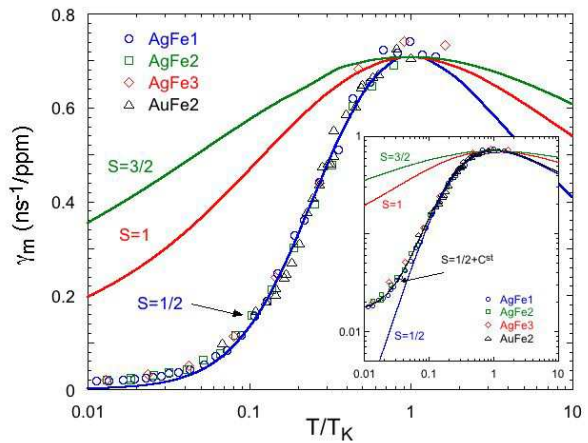}
  \caption{(color online) Dephasing rate per magnetic impurity
  as a function of $(T/T_K)$. The NRG results, keeping 1900 states for each NRG iteration and using the
discretization parameter $\Lambda=1.50$,
   for $S=1/2$ (blue), $S=1$ (red) and $S=3/2$ (green) have been scaled in temperature
  such that the maxima coincide with the experimental data. The black solid line in the inset corresponds to
  S=1/2 where we add a constant background.}
  \label{scaling}
\end{figure}

The phase coherence time $\tau_\phi$ for samples containing magnetic
Fe impurities shows a quite different behavior. One first observes a
plateau around $2\,K$ and subsequently a desaturation of the phase
coherence time. This can be understood within the framework of Kondo
physics. At temperatures above the Kondo scale $T_{K}$ spin flip
scattering is dominant and leads to a plateau in $\tau_\phi$. At
temperatures below $T_K$, the magnetic impurity spins get screened
and as a consequence the phase coherence time increases. By fitting
the data to eq.\,\ref{eq_tauphi} and taking into account magnetic
scattering in a similar way as was done in
ref.\,\cite{bauerle_prl_05}, we can extract a Kondo temperature
$T_{K}$ (using the conventions of \cite{rosch_prl_06}) as well as
the magnetic impurity concentration. A typical fit is shown by the
dotted blue line for sample AgFe1. We extract impurity
concentrations of 1.3, 13 and 33 ppm for sample AgFe\,1, AgFe\,2 and
AgFe\,3 respectively. The Kondo temperature is similar for all
samples and of the order of $T_K=4.3\,K\,\pm\,0.2K$. To extract the
dephasing rate due to magnetic impurities $\gamma_{m}$, we subtract
the experimentally measured electron-electron and electron-phonon
contribution. For the subtraction we use the coefficients obtained
for the clean sample. We then plot our data scaled per magnetic
impurity as a function of $(T/T_K)$ as shown on figure
\ref{scaling}. In addition we also plot data of sample AuFe2 of
ref.\cite{bauerle_prl_05} having a different Kondo temperature. All
data points fall on a \emph{universal} curve. In comparison we also
plot the NRG results for $S=1/2$ \cite{rosch_prl_06} as indicated by
the blue solid line. We find remarkable agreement with this theory
down to temperatures of $0.1\,T_K$. At lower temperatures, however,
we observe deviations from this theory, as emphasized on the log-log
plot in the inset of figure \ref{scaling}. The fact that we observe
a universal curve for different impurity concentrations which
depends \emph{only} on $T/T_{K}$ is a strong indication that we are
in the single impurity limit and interactions between the magnetic
impurities can be ruled out. On the other hand, it is clear that our
data are \emph{not} consistent with the Fermi liquid prediction of
the spin $1/2$ impurity model where a $T^2$ \cite{comment_FL}
temperature dependence of the magnetic scattering time is expected
at low temperature.

It is quite remarkable that the temperature dependence of $\gamma_m$
above $0.1 T_K$ is this well described by the S=1/2, single channel
model. Fe is expected to be characterized by both an orbital degree
of freedom and a much larger spin, $S=2$, coupling to electrons in
up to five angular momentum channels. The relevant theoretical model
depends on the position of the Fe within the Ag crystal structure,
crystal field effects and the strength of the spin-orbit coupling
and is presently not known. An obvious candidate to explain the weak
temperature dependence of $\gamma_m$ for low $T$ is to assume that
the large spin of Fe is only partially screened. This occurs when
the effective number of channels $n_c$ is smaller than $2 S$ (for
cubic symmetry $n_c=2$ or $3$ is expected).

We have therefore computed $\gamma_m$ and the Kondo contribution to
the resistance $R(T)$ for $1/2 \le S \le 5/2$ in the single channel,
$n_c=1$, limit as shown in Figs.~\ref{R_T} and \ref{scaling} for $S
\le 3/2$ \cite{comment_higher_S}. Already for $S=1$, the temperature
dependence of both, the resistivity as well as the dephasing rate is
much slower than for the case of $S=1/2$. By screening a single
channel, one can only reduce the size of the spin by $1/2$. The
resulting spin of size $S-1/2$ remains, however, so strongly coupled
to the electrons that $\gamma_m$ remains large even considerably
below $T_K$. The corresponding very broad maximum around $T_K$ is
clearly inconsistent with the experimental result already for $S=1$
as is shown in Fig.~\ref{scaling}. While the resistivity is much
less sensitive at low $T$, similar conclusions can be drawn by
comparing our results for $\Delta \rho(T)$ to the NRG theory in
Fig.~\ref{R_T}.

The comparisons show that the experimental data are indeed extremely
close to a spin 1/2 model and we can definitely rule out any
underscreened scenario. Also an overscreened Kondo model with $n_c>2
S$ can not explain the experimental data as it would lead to an even
larger dephasing rate at low $T$ \cite{delft_99}. Our data strongly
indicate a perfect screening of almost all of the Fe impurities.
Such a screening may, however, involve more than one channel and a
spin larger than 1/2. The fact that the fitted concentrations $n_S$
are $50\%$ lower than the nominal concentrations, and that the Kondo
temperatures obtained by fitting $\gamma_m$ and $\Delta \rho$ with
the $S=1/2$ single channel model differ almost by a factor
$T_K/T_K^\rho =1.4 \pm 0.3$ certainly supports this hypothesis. One
actually obtains $T_K/T_K^\rho\approx 1.1$ from the NRG for $S=1/2$
\cite{rosch_prl_06} and interestingly, calculations for the
underscreened single-channel model show that this ratio increases
with increasing spin up to $1.6$ for $S=5/2$.

What could then be the origin of this puzzling temperature
dependence of $\gamma_m$ below $0.1\,T_K$ ? One could argue that the
relatively weak temperature dependence below $0.1\,T_K$ originates
from some other source, different from magnetic impurities. As
$\gamma_m$ scales perfectly with the magnetic impurity concentration
we can definitely rule out any extrinsic effects. The curious
temperature dependence of $\gamma_m$ \emph{has to} originate from
the magnetic impurities themselves or from the implantation process.
A small fraction of implanted Fe impurities may for example end up
close to a lattice defect or in a grain boundary. One could also
imagine that the ion implantation creates additionally dynamical
defects, like two level systems (TLS) which can lead to a much
slower decay of the dephasing rate at low temperatures
\cite{delft_99}. Assuming a temperature independent background,
arising from such slow TLS, as indicated by the black solid line in
the inset of Fig.\ref{scaling}, results in a very good agreement
between the experimental data and the $S=1/2$ result. Let us
mention, however, that electron spin resonance experiments, where Cu
implantations have been performed into a Cu host, did not see any
difference before and after implantation down to temperatures of
3\,K \cite{these_hurdequint}. Note also that the AuFe sample, which
has been fabricated without ion implantation shows the same
universal temperature dependence down to $0.07\,T_K$.

In conclusion we have measured the phase coherence time of AgFe
quantum wires down to temperatures of $0.01\,T_K$. We have shown
that the magnetic contribution to the dephasing rate $\gamma_m$ per
magnetic impurity is a universal function of $T/T_K$. Its
temperature dependence is in remarkable agreement with recent NRG
studies for S=1/2, down to temperatures of 0.1\,$T_K$. At lower
temperatures, we observe deviations from this theory. The comparison
of the experimental data with our NRG calculations for higher spins
allows us to rule out any underscreened model to explain the
deviations at the lowest temperatures. Ion implantation of non
magnetic impurities and further tests of scaling should allow to
shed light onto this puzzling temperature dependence of $\gamma_m$
in the very low temperature limit. \\
\textit{note added}: We recently became aware of a related work
leading to similar conclusions [G.M. Alzoubi et al.].

We acknowledge helpful discussions with P. Simon, G. Zar\'and, L.
Glazman, A. Zawadowski, S. Kettemann, A. Altland, P. H. Dederichs,
A. Liebsch, H. Bouchiat, M. Lavagna and D. Feinberg. We are indebted
to the Quantronics group for the silver evaporation. This work has
been supported by the European Comission FP6 NMP-3 project 505457-1
Ultra-1D and ANR-PNANO \textit{QuSpin} and by SFB 608 and TR12 of
the DFG.


\begin{thebibliography}{99}
\bibitem{comment_SRIM} The SRIM simulations indicate that 22\% of the ions
penetrate into the substrate. This correction is taken into account
for the conversion into ppm (table \ref{table}).
\bibitem{comment_adding} Strictly speaking, a simple addition of dephasing rates is not correct,
see \cite{rosch_prl_06,gilles_book}. This effect is, however, very
small in the relevant $T$ range.
\bibitem{comment_higher_S} Similar calculations for the frequency-dependence of the inelastic cross section
at $T=0$ can be found in Ref.~\cite{hewson_04}, analytical results
for $\tau_m$ at low and high $T$ are discussed in
\cite{glazman_03}).
\end{thebibliography}
\end{document}